\documentclass[manuscript,screen]{acmart}
\AtBeginDocument{%
  \providecommand\BibTeX{{%
    \normalfont B\kern-0.5em{\scshape i\kern-0.25em b}\kern-0.8em\TeX}}}

\setcopyright{acmcopyright}
\copyrightyear{2023}
\acmYear{2023}
\acmDOI{XXXXXXX.XXXXXXX}

\usepackage{amsmath, amsthm, amsfonts}
\usepackage{graphicx}
\usepackage{textcomp}
\usepackage{xcolor}
\usepackage{comment}
\def\BibTeX{{\rm B\kern-.05em{\sc i\kern-.025em b}\kern-.08em
    T\kern-.1667em\lower.7ex\hbox{E}\kern-.125emX}}

\usepackage{algpseudocode}
\usepackage{algorithm}

\usepackage{caption}
\usepackage{subcaption}

\algnewcommand\algorithmicforeach{\textbf{for each}}
\algdef{S}[FOR]{ForEach}[1]{\algorithmicforeach\ #1\ \algorithmicdo}

\begin{document}

\title{Automated Arrangements of Multi-Part Music for Sets of Monophonic Instruments}

\author{Matthew McCloskey}
\author{Gabrielle Curcio}
\author{Amulya Badineni}
\author{Kevin McGrath}
\author{Dimitris Papamichail}
\email{papamicd@tcnj.edu}
\affiliation{%
  \institution{The College of New Jersey}
  \streetaddress{2000 Pennington Road}
  \city{Ewing}
  \state{New Jersey}
  \country{USA}
  \postcode{08618}
}


\renewcommand{\shortauthors}{McCloskey and Curcio, et al.}

\begin{abstract}
Arranging music for a different set of instruments that it was originally written for is traditionally a tedious and time-consuming process, performed by experts with intricate knowledge of the specific instruments and involving significant experimentation. In this paper we study the problem of automating music arrangements for music pieces written for monophonic instruments or voices. We designed and implemented an algorithm that can always produce a music arrangement when feasible by transposing the music piece to a different scale, permuting the assigned parts to instruments/voices, and transposing individual parts by one or more octaves. We also published open source software written in Python that processes MusicXML files and allows musicians to experiment with music arrangements. It is our hope that our software can serve as a platform for future extensions that will include music reductions and inclusion of polyphonic instruments.
\end{abstract}



\maketitle

\section{Introduction}

Music arrangements involve the adaptation of a piece of music for different instruments or ensembles. This allows the music to be performed in a variety of settings, enhances the repertory of musicians, and can also help to bring new life to a piece that may have been composed for a specific instrument or ensemble \cite{white92}. Additionally, arrangements can help to showcase the unique strengths of different instruments or even create entirely new interpretations of a piece. The process of arranging a piece of music can be a creative endeavor in itself, giving the arranger the opportunity to put their own spin on a familiar work, greatly enhancing the listening experience for audiences \cite{baker1988, elder2018, stefan2017}.

The computational complexity of arranging music written for a set of instruments toward a target single instrument, often employing reasonable reductive constraints, has been examined in the work of Moses and Demaine \cite{moses17}. Complexities of dealing with polyphonic instruments, such as piano and guitar, include the need of considering possible fingerings as well as reductions, the elimination of certain notes for playability of even feasibility. Most research in automating music arrangements has concentrated on the piano, primarily concerning orchestral pieces \cite{Chiu09, onuma10, huang12, Takamori19, Nakamura18, moyu22}. Much of that work involves reductions to enable feasibility. Other work in the field has examined arrangements for the guitar \cite{tuohy06, hori12, Hori13}, 
wind ensembles \cite{Maekawa06}, and other orchestral instruments \cite{Crestel19}.

Despite its obvious benefits, we are not aware of any published algorithm or widely available software that allows for the automated arrangement of a given music piece to a different set of instruments that it was originally written for in the general case. Working toward filling that need, we designed and implemented an algorithm that arranges music written for monophonic instruments and guarantees a successful outcome when an arrangement is possible without score reduction. Our recursive backtracking algorithm exhaustively examines all feasible assignments of parts to available instruments and all possible transpositions of the piece, including independent octave transpositions of individual parts, to determine a successful arrangement that minimally affects the musicality of the piece.

\section{Methods}

\subsection{Definitions}

For the purposes of our research, a music piece is written in a chromatic scale and notes are separated by the interval of a semitone. We will assume that all notes fall within a total range of $88$ semitones, the notes of a traditional piano, from A0 to C8. We will assign an integer to each note in the range, such that all notes can be represented by an integer from 1 to 88. For our discussion, a monophonic instrument is one that can only play one pitch at a time, such as the flute, the oboe, or a voice. Polyphonic instruments can play multiple notes simultaneously, such as the piano, guitar, or harp. A polyphonic instrument can always play a monophonic part within its range.

\begin{figure}
\centering
	\includegraphics[width=0.8\linewidth]{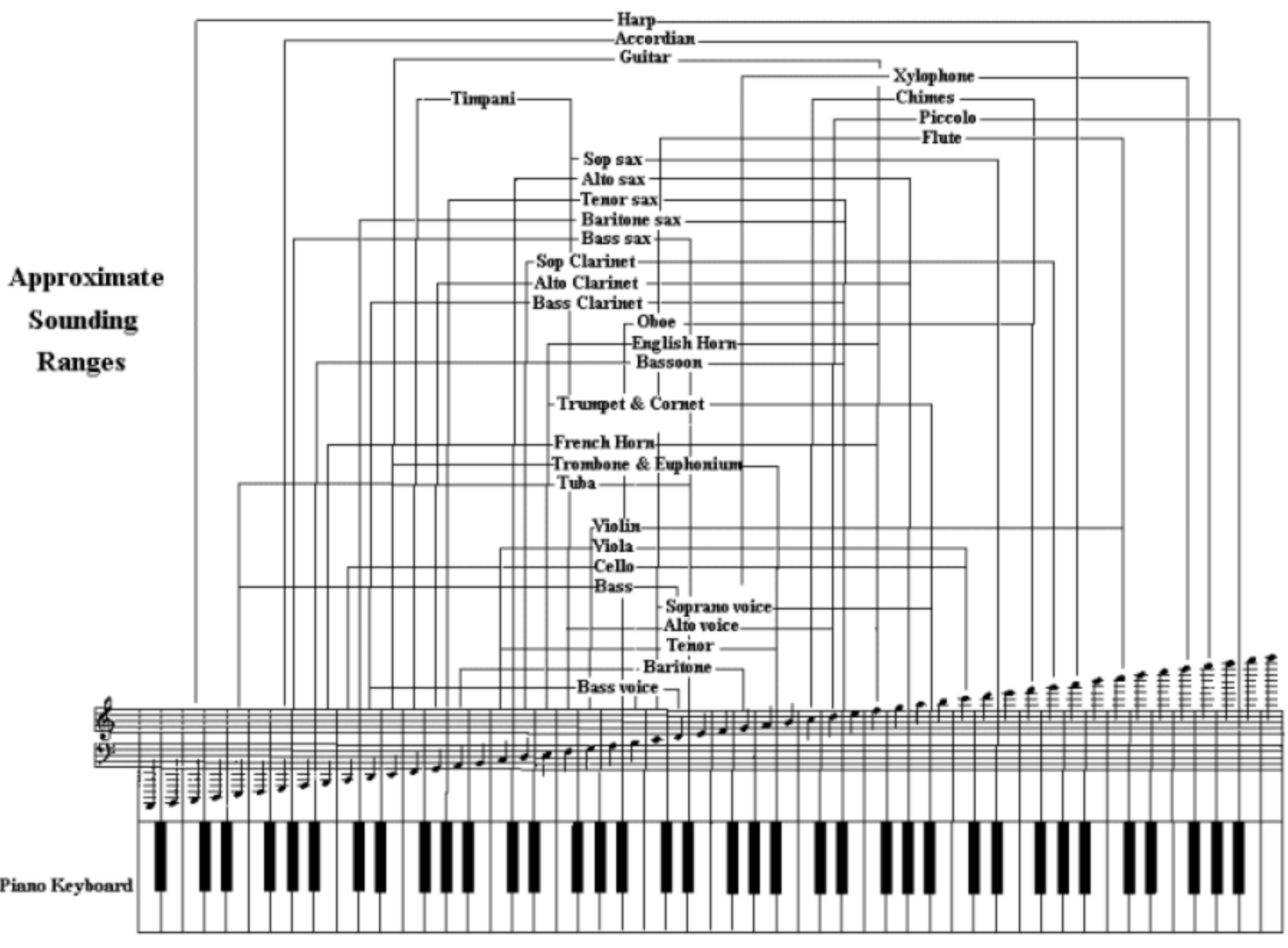}
    \caption{ Approximate sounding ranges of instruments and voices. Figure reproduced with permission from Dr. Brian Blood (dolmetch.com)}
\label{fig:ranges}
\end{figure} 

 For our study an input music piece will consist of $n$ parts, each being assigned to a single monophonic instrument or voice. Such parts are presented in the sheet music representation of the piece in an equal number of staves each. Our algorithm preserves the rhythm, rhythmic values of notes and rests, as well as bar lines of the music piece. Clefs, key signatures and accidentals are adjusted based on the scale of the transposed music and the instruments/voices that parts are assigned to. Our algorithm does not control for instrument timbre that may be expected in any part of the music; similarly, the thickness of the piece is not being necessarily maintained.

We will assume that an input music piece is originally written for $n$ instruments $I_1, I_2, \cdots, I_n$, each assigned to play a part $P_i$ of the piece, with $1 \le i \le n$. We seek to arrange the music for $n$ output instruments $O_1, O_2, \cdots, O_n$. The range of each part $i$ is an integer interval $R_i = \llbracket a_i, b_i\rrbracket$, where $a_i$ is the integer value corresponding to the lowest frequency note and $b_i$ to the highest frequency note played by instrument $I_i$ in part $P_i$, $1 \le i \le n$. Likewise, the playing range of each output instrument $O_i$ will be denoted by $OR_i$, $1 \le i \le n$, indicating the integer interval of values corresponding to the notes the instrument is able to play. Approximate ranges for a set of instruments and voices can be seen in Figure \ref{fig:ranges}.

\subsection{Monophonic instrument set arrangement algorithm} \label{sec:algo}

Our Monophonic Music Arrangement (MMS) algorithm performs a nearly comprehensive search of possible permutations of parts. The music is transposed to all twelve keys, and the algorithm runs on each key, unless a solution has been found so far that results in fewer sharps/flats over all keys for each part. This is designed to prevent the "ideal" transposition from having a complex key signature if not necessary. Other than that, the search is fully comprehensive. For each part, the algorithm finds all possible transpositions of each part in the source piece that can be played by at least one available instrument. All permutations of these possible transpositions are then examined. If all parts can be played by at least one instrument, the algorithm then checks if there exists a set of part assignments that is valid. This is performed by a recursive function that is memoized to improve performance. If a transposed key yields valid permutations, the transposition with the least total deviation from the original composition is selected. Once all twelve keys have been checked, all permutations are tried using the selected transposition, unless there is no selected transposition, in which case the algorithm fails.
All permutations are checked, and for those that are valid in the given transposition, the best arrangement is selected based on how closely the average pitch of each part matches the median pitch of the instrument's range. 

The MMA algorithm implementation consists of four main function described in  pseudocode below.

\begin{algorithm}
\caption{Find Transposed Options}
\begin{algorithmic}
\Procedure{FindTransposedOptions}{$originalStream, arrangementParts, semitones$}
\State $stream \xleftarrow{}$ $originalStream$ transposed by given $semitones$
\State $parts \xleftarrow{}$ new list
\For{$part$ in $stream$}
    \State $choices \xleftarrow{}$ new list
    \For{each $transposition$} 
        \State $set \xleftarrow{}$ the subset of $arrangementParts$ that can play at this transposition
        \State add $(semitones + transposition, set)$ to $choices$
    \EndFor
    \If{$choices$ is empty}
        \State \Return null
    \EndIf
    \State add $choices$ to $parts$
\EndFor
\State \Return $parts$
\EndProcedure
\end{algorithmic}
\end{algorithm}

\begin{algorithm}
\caption{Run Transposed}
\begin{algorithmic}
\Procedure{RunTransposed}{$stream, parts, semitones$}
\State $selections \xleftarrow{}$ new list
\For{$option$ in all possible transpositions from \Call{FindTransposedOptions}{$stream, parts, semitones$}}
    \State $partsCovered \xleftarrow{}$ new list
    \State $selection \xleftarrow{}$ new list
    \For{$transposition$ in $option$}
        \State add set of parts covered to $partsCovered$
        \State add deviation of transposition to $selection$
    \EndFor
    \State $allPartsCovered \xleftarrow{}$ the union of all sets in $partsCovered$
    \If{$allPartsCovered$ contains all parts and \Call{$ValidateArrangement$}{$parts$, $partsCovered$, $allPartsCovered$}}
        \State add $selection$ to $selections$
    \EndIf
\EndFor
\State \Return $selections$
\EndProcedure
\end{algorithmic}
\end{algorithm}

\begin{algorithm}
\caption{Find Best Choice}
\begin{algorithmic}
\Procedure{FindBestChoice}{$stream, parts$}
\State $bestChoice \xleftarrow{}$ null
\State $bestSharps \xleftarrow{} \infty$
\For{$semitones$ from $-6$ through $5$}
    \State $sharps \xleftarrow{}$ the total number of sharps/flats that would appear in the key signature for each part
    \If{$sharps \le bestSharps$}
        \State $thisBestChoice \xleftarrow{}$ element from \Call{RunTransposed}{$stream, parts, semitones$} with the least deviation
        \If{$thisBestChoice \ne$ null and either $sharps < bestSharps$ or deviation of $thisBestChoice <$ deviation of $bestChoice$}
            \State $bestChoice \xleftarrow{} thisBestChoice$
            \State $bestSharps \xleftarrow{} thisBestSharps$
        \EndIf
    \EndIf
\EndFor
\State \Return $bestChoice$
\EndProcedure
\end{algorithmic}
\end{algorithm}

\begin{algorithm}
\caption{MMA Algorithm}
\begin{algorithmic}
\Procedure{MMA}{$stream, parts$}
\State $bestChoice \xleftarrow{}$ \Call{FindBestChoice}{$stream, parts$}
\If{$bestChoice =$ null}
    \State \Return null
\EndIf
\State transpose each part by the resulting transposition
\State $bestFit \xleftarrow{} \infty$
\For{each permutation of $newParts$}
    \If{all parts are valid in the given permutation}
        \State $fit \xleftarrow{}$ the total absolute difference between the average pitches and the median pitch of each part
        \If{$fit < bestFit$}
            \State $bestFit \xleftarrow{} fit$
            \State $bestPermutation \xleftarrow{}$ this permutation
        \EndIf
    \EndIf
\EndFor
\State \Return $bestPermutation$
\EndProcedure
\end{algorithmic}
\end{algorithm}

\subsection{Implementation}

\begin{figure}
\centering
    \begin{subfigure}[!tbh]{0.48\textwidth}
        \centering
\begin{verbatim}
clarinet = 1
tenor-sax = 2
alto-sax = 2
\end{verbatim}
\caption{An example arrangement file}
\label{fig:arrangement-toml}
\end{subfigure}
    \hfill
\begin{subfigure}[!tbh]{0.48\textwidth}
\begin{verbatim}
[alto-sax]
name = "AltoSaxophone"
minimum = "Db3"
maximum = "Bb5"
key = "Eb"
\end{verbatim}
\caption{An entry in the instrument metadata file}
\label{fig:metadata-toml}
\end{subfigure}
\caption{Examples of input instrument set and instrument information files}
\label{fig:input}
\end{figure}

The MMA algorithm was implemented in Python utilizing the Music21 library and the MuseScore software. Our program requires two input files and produces a single output file with the music arrangement. The required input files consist of the original piece of music in MusicXML format and a TOML file listing the instrument set to arrange for, where an assigned value of $k$ to an instrument indicates $k$ parts should be arranged for that instrument. An example of a TOML file with an input instrument set consisting of one clarinet, two tenor saxophones, and two alto saxophones is shown in Figure \ref{fig:arrangement-toml}. Metadata about each instrument, consisting of its key in notation and a reasonable note range, is defined in a separate TOML file which is loaded separately by the program and is populated with common music instruments. An example of an entry for the alto saxophone in the instrument metadata file is shown in Figure \ref{fig:metadata-toml}.

During execution our program checks whether the number of input instruments matches the number of parts in the piece, and then attempts to arrange for the given instruments as previously described. If arrangements are found, the best arrangement based on the criteria described in section \ref{sec:algo} is output as a MusicXML file. If no feasible arrangement is found, or if the number of instruments does not match, then an error message is displayed and no output file is produced.

\section{Results}

\begin{figure}[b]
\centering
    \begin{subfigure}{0.48\textwidth}
        \centering
        \includegraphics[width=\linewidth]{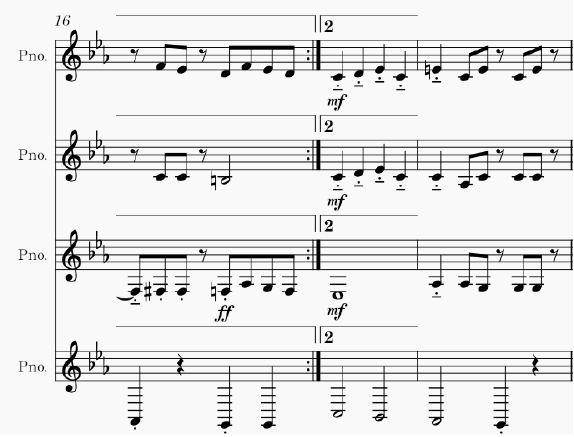}
        \caption{Original Score}
        \label{subfig:orig1}
    \end{subfigure}
    \hfill
    \begin{subfigure}{0.48\textwidth}
        \centering
	\includegraphics[width=\linewidth]{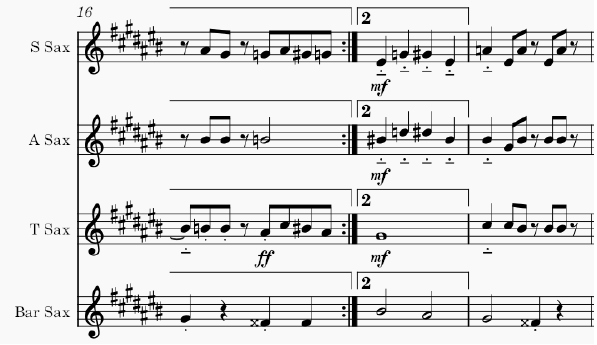}
        \caption{Arranged score}
        \label{subfig:out1}
    \end{subfigure}
\caption{Three measures from an arrangement of 'Puttin' on the Ritz' from piano to saxophone quartet}
\label{fig:example1}
\end{figure}

\begin{figure}
\centering
    \begin{subfigure}{0.48\textwidth}
        \centering
        \includegraphics[width=\linewidth]{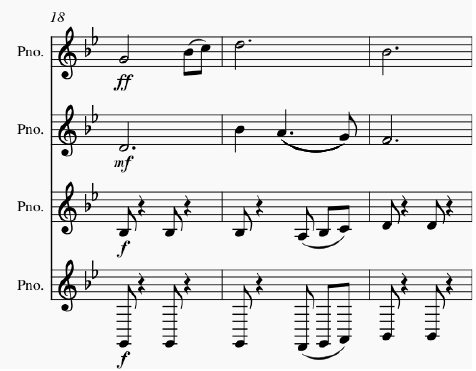}
        \caption{Original Score}
        \label{subfig:orig1}
    \end{subfigure}
    \hfill
    \begin{subfigure}{0.48\textwidth}
        \centering
	\includegraphics[width=\linewidth]{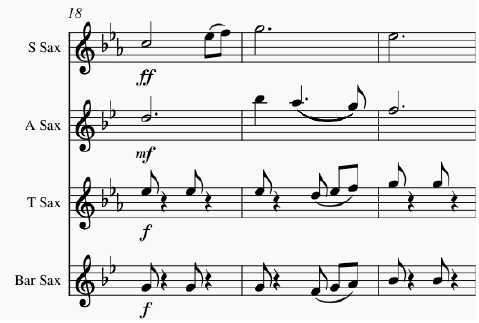}
        \caption{Arranged score}
        \label{subfig:out1}
    \end{subfigure}
\caption{Three measures from an arrangement of 'Carol of the Bells' from voices to saxophone quartet}
\label{fig:example2}
\end{figure}

We tested our software on a variety of music pieces written for monophonic instruments. In Figure \ref{fig:example1} we show three measures, starting at measure $16$, of the {\it Puttin' on the Ritz} song by Irving Berlin. Part (a) shows the input score composed of four monophonic parts. Part (b) displays the arranged piece for saxophone quartet, consisting of a soprano, alto, tenor, and baritone saxophones. Similarly, in Figure \ref{fig:example2} we display three measures of {\it Carol of the Bells}, as arranged and performed by the Pentatonix voice group, starting at measure 18 of the piece.

Complete input/output files for three test cases of our software, including the {\it Puttin' on the Ritz} and {\it Carol of the Bells} above, can be examined at: https://owd.tcnj.edu/$\sim$papamicd/music/mma/examples/

The repository for this project can be found at: https://github.com/spazzylemons/music-arrangement/

\section{Conclusions and future work}

Our monophonic music arrangement algorithm and its software implementation create a platform for automating music arrangements with minimal user input. Although currently basic in its functionality, it can be readily extended in a number of different directions. For accommodating arrangements for a smaller sets of instruments than the number of parts in the music, score reduction techniques can be applied to eliminate certain parts or at least reduce the number of simultaneous notes that are played throughout the piece, while maintaining faithfulness to the original. To allow for the inclusion of polyphonic instruments in the arrangements, further work is required in analyzing and decomposing polyphonic parts into monophonic ones and inversely, while adhering to constraints related to fingerings and other instrument and player restrictions. 

\begin{acks}
The authors acknowledge use of the ELSA high performance computing cluster at The College of New Jersey for conducting the research reported in this paper. This cluster is funded in part by the National Science Foundation under grant numbers OAC-1826915 and OAC-1828163.
\end{acks}

\bibliographystyle{ACM-Reference-Format}
\bibliography{paper.bib}


\begin{thebibliography}{16}


\ifx \showCODEN    \undefined \def \showCODEN     #1{\unskip}     \fi
\ifx \showDOI      \undefined \def \showDOI       #1{#1}\fi
\ifx \showISBNx    \undefined \def \showISBNx     #1{\unskip}     \fi
\ifx \showISBNxiii \undefined \def \showISBNxiii  #1{\unskip}     \fi
\ifx \showISSN     \undefined \def \showISSN      #1{\unskip}     \fi
\ifx \showLCCN     \undefined \def \showLCCN      #1{\unskip}     \fi
\ifx \shownote     \undefined \def \shownote      #1{#1}          \fi
\ifx \showarticletitle \undefined \def \showarticletitle #1{#1}   \fi
\ifx \showURL      \undefined \def \showURL       {\relax}        \fi
\providecommand\bibfield[2]{#2}
\providecommand\bibinfo[2]{#2}
\providecommand\natexlab[1]{#1}
\providecommand\showeprint[2][]{arXiv:#2}

\bibitem[Baker(1988)]%
        {baker1988}
\bibfield{author}{\bibinfo{person}{D. Baker}.} \bibinfo{year}{1988}\natexlab{}.
\newblock \bibinfo{booktitle}{\emph{David Baker's Arranging \& Composing: For
  the Small Ensemble, Jazz, R \& B, Jazz-rock}}.
\newblock \bibinfo{publisher}{Alfred Publishing Company}.
\newblock
\showISBNx{9780882844695}
\showLCCN{92247055}
\urldef\tempurl%
\url{https://books.google.com/books?id=Le0EnwEACAAJ}
\showURL{%
\tempurl}


\bibitem[Chiu et~al\mbox{.}(2009)]%
        {Chiu09}
\bibfield{author}{\bibinfo{person}{Shih~Chuan Chiu}, \bibinfo{person}{Man~Kwan
  Shan}, {and} \bibinfo{person}{Jiun~Long Huang}.}
  \bibinfo{year}{2009}\natexlab{}.
\newblock \showarticletitle{{Automatic system for the arrangement of piano
  reductions}}. In \bibinfo{booktitle}{\emph{ISM 2009 - 11th IEEE International
  Symposium on Multimedia}}.
\newblock
\showISBNx{9780769538907}
\urldef\tempurl%
\url{https://doi.org/10.1109/ISM.2009.105}
\showDOI{\tempurl}


\bibitem[Crestel and Esling(2019)]%
        {Crestel19}
\bibfield{author}{\bibinfo{person}{L{\'{e}}opold Crestel} {and}
  \bibinfo{person}{Philippe Esling}.} \bibinfo{year}{2019}\natexlab{}.
\newblock \showarticletitle{{Live orchestral piano, a system for real-time
  orchestral music generation}}. In \bibinfo{booktitle}{\emph{Proceedings of
  the 14th Sound and Music Computing Conference 2017, SMC 2017}}.
\newblock
\showISBNx{9789526037295}
\showeprint[arxiv]{1609.01203}


\bibitem[Demaine and Moses(2017)]%
        {moses17}
\bibfield{author}{\bibinfo{person}{Erik~D. Demaine} {and}
  \bibinfo{person}{William~S. Moses}.} \bibinfo{year}{2017}\natexlab{}.
\newblock \showarticletitle{{364Computational Complexity of Arranging Music}}.
\newblock In \bibinfo{booktitle}{\emph{{The Mathematics of Various Entertaining
  Subjects: Research in Games, Graphs, Counting, and Complexity, Volume 2}}}.
  \bibinfo{publisher}{Princeton University Press}.
\newblock
\showISBNx{9780691171920}
\urldef\tempurl%
\url{https://doi.org/10.23943/princeton/9780691171920.003.0019}
\showDOI{\tempurl}


\bibitem[Elder(2018)]%
        {elder2018}
\bibfield{author}{\bibinfo{person}{J.B. Elder}.}
  \bibinfo{year}{2018}\natexlab{}.
\newblock \bibinfo{booktitle}{\emph{The Art of Arranging and Orchestration}}.
\newblock \bibinfo{publisher}{Independently Published}.
\newblock
\showISBNx{9781717843531}
\urldef\tempurl%
\url{https://books.google.com/books?id=mOQKugEACAAJ}
\showURL{%
\tempurl}


\bibitem[Hori et~al\mbox{.}(2013)]%
        {Hori13}
\bibfield{author}{\bibinfo{person}{Gen Hori}, \bibinfo{person}{Hirokazu
  Kameoka}, {and} \bibinfo{person}{Shigeki Sagayama}.}
  \bibinfo{year}{2013}\natexlab{}.
\newblock \showarticletitle{{Input-output HMM applied to automatic arrangement
  for guitars}}.
\newblock \bibinfo{journal}{\emph{Journal of Information Processing}}
  (\bibinfo{year}{2013}).
\newblock
\showISSN{18826652}
\urldef\tempurl%
\url{https://doi.org/10.2197/ipsjjip.21.264}
\showDOI{\tempurl}


\bibitem[Hori et~al\mbox{.}(2012)]%
        {hori12}
\bibfield{author}{\bibinfo{person}{Gen Hori}, \bibinfo{person}{Yuma Yoshinaga},
  \bibinfo{person}{Satoru Fukayama}, \bibinfo{person}{Hirokazu Kameoka}, {and}
  \bibinfo{person}{Shigeki Sagayama}.} \bibinfo{year}{2012}\natexlab{}.
\newblock \showarticletitle{Automatic arrangement for guitars using hidden
  Markov model}.
\newblock \bibinfo{journal}{\emph{Proceedings of 9th Sound and Music Computing
  Conference (SMC2012)}} (\bibinfo{date}{7} \bibinfo{year}{2012}),
  \bibinfo{pages}{450--455}.
\newblock


\bibitem[Huang et~al\mbox{.}(2012)]%
        {huang12}
\bibfield{author}{\bibinfo{person}{Jiun-Long Huang},
  \bibinfo{person}{Shih-Chuan Chiu}, {and} \bibinfo{person}{Man-Kwan Shan}.}
  \bibinfo{year}{2012}\natexlab{}.
\newblock \showarticletitle{Towards an Automatic Music Arrangement Framework
  Using Score Reduction}.
\newblock \bibinfo{journal}{\emph{ACM Trans. Multimedia Comput. Commun. Appl.}}
  \bibinfo{volume}{8}, \bibinfo{number}{1}, Article \bibinfo{articleno}{8}
  (\bibinfo{date}{feb} \bibinfo{year}{2012}), \bibinfo{numpages}{23}~pages.
\newblock
\showISSN{1551-6857}
\urldef\tempurl%
\url{https://doi.org/10.1145/2071396.2071404}
\showDOI{\tempurl}


\bibitem[Maekawa et~al\mbox{.}(2006)]%
        {Maekawa06}
\bibfield{author}{\bibinfo{person}{Hiroshi Maekawa}, \bibinfo{person}{Norio
  Emura}, \bibinfo{person}{Masanobu Miura}, {and} \bibinfo{person}{Masuzo
  Yanagida}.} \bibinfo{year}{2006}\natexlab{}.
\newblock \showarticletitle{{On machine arrangement for smaller wind-orchestras
  based on scores for standard wind-orchestras}}. In
  \bibinfo{booktitle}{\emph{International Conference on Music Perception and
  Cognition, ICMPC 2006}}. \bibinfo{pages}{268--273}.
\newblock


\bibitem[Nakamura and Yoshii(2018)]%
        {Nakamura18}
\bibfield{author}{\bibinfo{person}{Eita Nakamura} {and}
  \bibinfo{person}{Kazuyoshi Yoshii}.} \bibinfo{year}{2018}\natexlab{}.
\newblock \showarticletitle{{Statistical piano reduction controlling
  performance difficulty}}.
\newblock \bibinfo{journal}{\emph{APSIPA Transactions on Signal and Information
  Processing}} (\bibinfo{year}{2018}).
\newblock
\showISSN{20487703}
\urldef\tempurl%
\url{https://doi.org/10.1017/ATSIP.2018.18}
\showDOI{\tempurl}
\showeprint[arxiv]{1808.05006}


\bibitem[Onuma and Hamanaka(2010)]%
        {onuma10}
\bibfield{author}{\bibinfo{person}{Sho Onuma} {and} \bibinfo{person}{Masatoshi
  Hamanaka}.} \bibinfo{year}{2010}\natexlab{}.
\newblock \showarticletitle{{Piano arrangement system based on composers'
  arrangement processes}}. In \bibinfo{booktitle}{\emph{International Computer
  Music Conference, ICMC 2010}}.
\newblock
\showISBNx{0971319286}


\bibitem[Stefan~Kostka et~al\mbox{.}(2017)]%
        {stefan2017}
\bibfield{author}{\bibinfo{person}{T.H. Stefan~Kostka}, \bibinfo{person}{T.H.
  Dorothy~Payne}, {and} \bibinfo{person}{B. Alm{\'e}n}.}
  \bibinfo{year}{2017}\natexlab{}.
\newblock \bibinfo{booktitle}{\emph{Tonal Harmony}}.
\newblock \bibinfo{publisher}{McGraw-Hill Education}.
\newblock
\showISBNx{9781259447099}
\showLCCN{2016052221}
\urldef\tempurl%
\url{https://books.google.com/books?id=Cs2UAQAACAAJ}
\showURL{%
\tempurl}


\bibitem[Takamori et~al\mbox{.}(2019)]%
        {Takamori19}
\bibfield{author}{\bibinfo{person}{Hirofumi Takamori}, \bibinfo{person}{Haruki
  Sato}, \bibinfo{person}{Takayuki Nakatsuka}, {and} \bibinfo{person}{Shigeo
  Morishima}.} \bibinfo{year}{2019}\natexlab{}.
\newblock \showarticletitle{{Automatic arranging musical score for piano using
  important musical elements}}. In \bibinfo{booktitle}{\emph{Proceedings of the
  14th Sound and Music Computing Conference 2017, SMC 2017}}.
\newblock
\showISBNx{9789526037295}


\bibitem[Terao et~al\mbox{.}(2022)]%
        {moyu22}
\bibfield{author}{\bibinfo{person}{Moyu Terao}, \bibinfo{person}{Yuki
  Hiramatsu}, \bibinfo{person}{Ryoto Ishizuka}, \bibinfo{person}{Yiming Wu},
  {and} \bibinfo{person}{Kazuyoshi Yoshii}.} \bibinfo{year}{2022}\natexlab{}.
\newblock \showarticletitle{Difficulty-Aware Neural Band-to-Piano Score
  Arrangement based on Note- and Statistic-Level Criteria}. In
  \bibinfo{booktitle}{\emph{ICASSP 2022 - 2022 IEEE International Conference on
  Acoustics, Speech and Signal Processing (ICASSP)}}.
  \bibinfo{pages}{196--200}.
\newblock
\urldef\tempurl%
\url{https://doi.org/10.1109/ICASSP43922.2022.9747615}
\showDOI{\tempurl}


\bibitem[Tuohy and Potter(2006)]%
        {tuohy06}
\bibfield{author}{\bibinfo{person}{D.R. Tuohy} {and} \bibinfo{person}{W.D.
  Potter}.} \bibinfo{year}{2006}\natexlab{}.
\newblock \showarticletitle{GA-based Music Arranging for Guitar}. In
  \bibinfo{booktitle}{\emph{2006 IEEE International Conference on Evolutionary
  Computation}}. \bibinfo{pages}{1065--1070}.
\newblock
\urldef\tempurl%
\url{https://doi.org/10.1109/CEC.2006.1688427}
\showDOI{\tempurl}


\bibitem[White(1992)]%
        {white92}
\bibfield{author}{\bibinfo{person}{G.C. White}.}
  \bibinfo{year}{1992}\natexlab{}.
\newblock \bibinfo{booktitle}{\emph{Instrumental Arranging}}.
\newblock \bibinfo{publisher}{McGraw-Hill}.
\newblock
\showISBNx{9780073018232}
\urldef\tempurl%
\url{https://books.google.com/books?id=4c2KMwEACAAJ}
\showURL{%
\tempurl}


\end{thebibliography}

\end{document}